# A Comparative Study of Smartphone and Smart TV Apps

Yonghui Liu[a], Xiao Chen[a], Yue Liu[a], Pingfan Kong[b], Tegawendé F. Bissyandé[b], Jacques Klein[b], Xiaoyu Sun[a], Li Li[a,*], Chunyang Chen[a] and John Grundy[a]

[a]Monash University, Australia
[b]University of Luxembourg, Luxembourg

ABSTRACT

**Context:** Smart TVs have become one of the most popular television types. Many app developers and service providers have designed TV versions for their smartphone applications. Despite the extensive studies on mobile app analysis, its TV equivalents receive far too little attention. The relationship between phone and TV has not been the subject of research works.
**Objective:** In this paper, we aim to characterize the relationship between smartphone and smart TV apps. To fill this gap, we conduct a comparative study on smartphone and smart TV apps in this work, which is the starting and fundamental step to uncover the domain-specific challenges.
**Method:** We gather a large-scale phone/TV app pairs from Google Play Store. We then analyzed the app pairs quantitatively and qualitatively from a variety of perspectives, including non-code (e.g., metadata, resources, permissions, etc.), code (e.g., components, methods, user interactions, etc.), security and privacy (e.g., reports of AndroBugs and FlowDroid).
**Results:** Our experimental results indicate that (1) the code of the smartphone and TV apps can be released in the same app package or in separate app packages with the same package name; (2) 43% of resource files and 50% of code methods are reused between phone/TV app pairs; (3) TV and phone versions of the same app often encounter different kinds of security vulnerabilities; and (4) TV apps encounter fewer user interactions than their phone versions, but the type of user interaction events, surprisingly, are similar between phone/TV apps.
**Conclution:** Our findings are valuable for developers and academics in comprehending the TV app ecosystem by providing additional insight into the migration of phone apps to TVs and the design mechanism of analysis tools for TV apps.

## 1. Introduction

Smart TV has become the dominant TV type nowadays. More and more users are switching from traditional TVs to Smart TVs [20], which offer additional features and conveniences such as allowing users to watch Netflix and YouTube directly on TVs through internet connectivity. According to [18], the global smart TV market is expected to reach a value of USD 253 billion by 2023. Among various types of Smart TVs on the market, Android TV, with a lot of benefits extended from the popular Android ecosystem, has undoubtedly become one of the most popular ones [41]. Similar to Android phones, Android TVs enable users to connect to the Google Play store to download and update apps, as well as utilize Google Assistant to accomplish hands-free tasks.

Despite the growing momentum of the smart TV industry (particularly in terms of the number of TV devices accessible in the Android ecosystem), the number of available TV applications is significantly less than the number of existing smartphone apps. Indeed, compared to over 2.6 million Android apps on Google Play, the number of Android apps for smart TVs is only around 7,000 [11]. Unfortunately, while extending the benefits of the Android ecosystem, Android TVs also extend the potential security flaws appearing in other Android-based devices such as smartphones or smartwatches. Additionally, customization made for Android TVs may introduce additional security risks for TV users. As discussed by [1], researchers discovered that there are 37 unique vulnerabilities, which could lead to high-impact cyber threats, available in 11 Android TV boxes through log-guided fuzzing.

There is an easily overlooked gap between the smartphone and smart TV (hereafter, TV) apps, making the prospect of TV apps left far behind the smartphone. Specifically, TV apps have a considerably different design protocol than smartphone apps due to the differences in the hardware, such as screen size, computation capability, and power supply. Liu *et al.* [30] and Tileria *et al.* [41] perform a security and privacy analysis of the Android TV applications. However, prior studies examined the Android TV apps only. The differences between Android mobile apps and TV apps are still poorly understood. Indeed, our lightweight literature search (over Google Scholar with different combinations of keywords including *Android TV app*, *Smart TV app*, *similarity*, *security*, *privacy*, *vulnerability*, etc.) reveals the community has not yet explored the realm of comparing smartphone and TV apps. To fill the gap, we conduct an exploratory study on Android phone and TV apps in this work. Through this study, we aim to understand the landscape of Android TV apps and the similarity between phone/TV app pairs and their security situations, and subsequently observe actionable insights towards achieving a better Android TV ecosystem.

Unfortunately, there is no publicly available dataset that is primarily composed of *Android phone/TV app pairs*. To this end, we firstly collect Android TV apps from Google Play to fulfill this research gap. By aggregating all acquired

ORCID(s):

applications, we finally construct a dataset consisting of 3,445 phone/TV app pairs. A mixed-methods approach is then used to characterize these app pairs with both quantitative and qualitative investigations. For instance, we investigate their parallels and contrasts at the non-code and code levels, e.g., on how they handle app permissions and user interactions.

This empirical study compares and contrasts Android apps on smartphones and smart TVs. We analyzed the app pairs quantitatively and qualitatively from a variety of perspectives, including non-code (e.g., metadata, resources, permissions, etc.), code (e.g., components, methods, user interactions, etc.), security and privacy (e.g., reports of AndroBugs and FlowDroid). We then provide several insightful discussions for future works on our deep analysis. The key contributions of this research are summarized as follows:

- **Dataset.** Since this is the first study to explore the similarities and differences between phone and TV apps, there are no publicly-available phone/TV app pair datasets to be used. We collected 3,445 pairs of Android phone/TV apps from Google Play Store. To foster research in this direction, we make our datasets[1] publicly available.

- **Artefact.** We perform a similarity analysis on the non-code and code components of phone/TV app pairs. Our results demonstrate that TV apps share a great deal of same non-code and code components with their mobile counterparts. But due to the platform differences, we observe that TV apps declare fewer permissions compared to their phone counterparts.

- **Security and Privacy.** We analyze the security and privacy risks in Android phone/TV pairs. We find that more privacy leaks are detected in Android TV applications.

- **User Interaction.** We analyze the differences in user interaction between smartphone and TV applications. Our results indicate that user interactions have been considerably decreased in the TV applications compared to their mobile counterparts.

The remainder of the paper is organized as follows. Section 2 discusses the study's rationale and motivation. Section 3 details the data collection procedure, and data characteristics. Section 4 offers the results and conclusions of our empirical investigation, followed by a discussion of the study's implications and potential risks to validity in Section 5. Section 6 discusses relevant literature, and section 7 concludes the article.

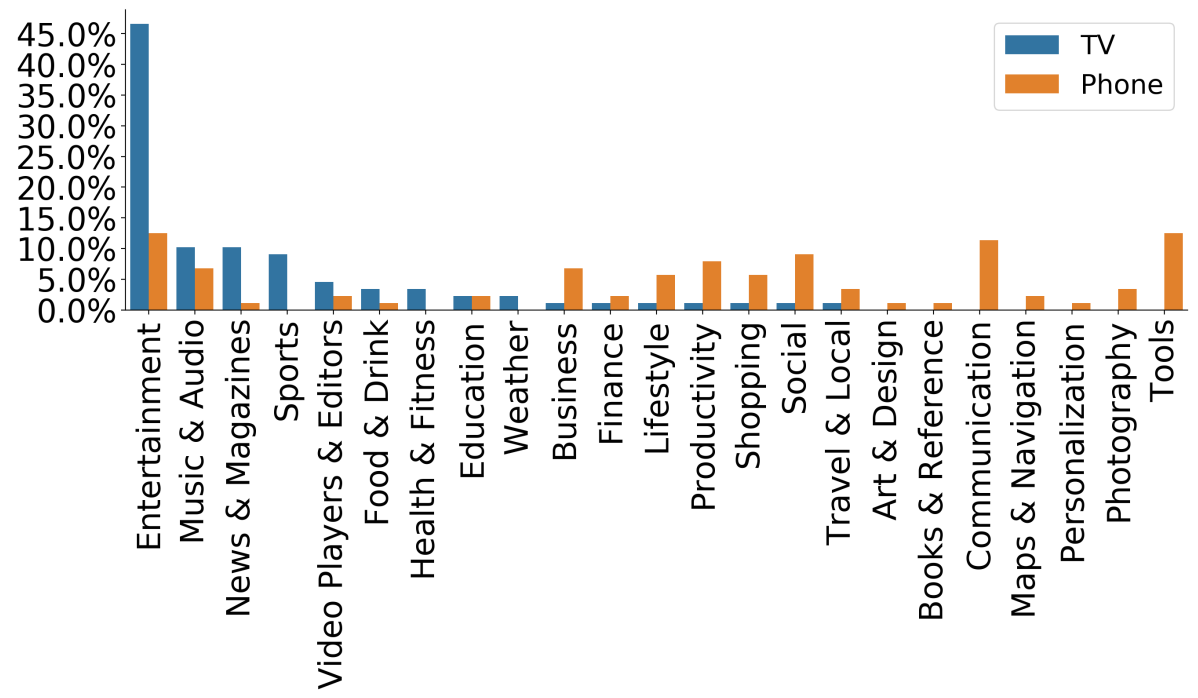

**Figure 1:** Category Distribution of 100 Most Popular Smartphone and Smart TV Apps on Google Play.

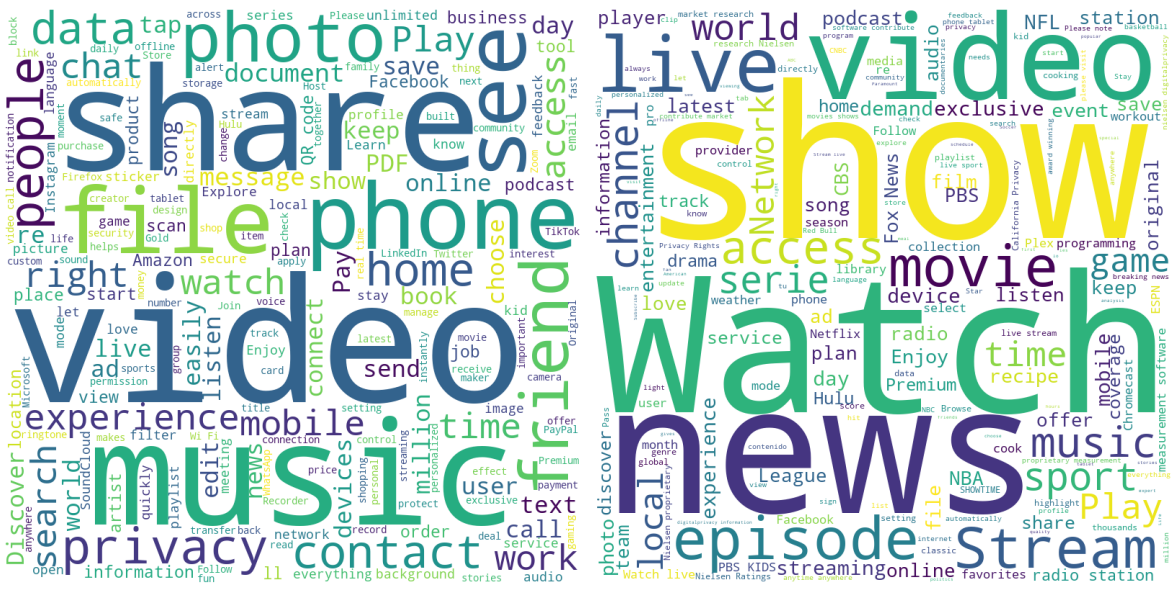

(a) Top 100 phone Apps (b) Top 100 TV Apps

**Figure 2:** Word Cloud of descriptions for apps collected.

## 2. Motivation

Smartphones and smart TVs perform different tasks in people's daily activities. Smartphones have become integrated into our daily lives. Many of us depend on smartphones for daily routines like chatting with friends and paying bills, thanks to the phone's portability. Smart TVs, on the other hand, are better suited to immersive group activities such as watching movies and playing games on the large screen of the TV. Thus, it may be diverse in how individuals use these devices. To better understand this, we conduct a preliminary study. Figure 1 extracts data from Google Play Store for the 100 most popular smartphone and TV apps, respectively, and visualizes their category distributions. It's obvious that the two platforms' most popular apps are classified differently. *Entertainment*, for example, is the most popular category in the TV app market, accounting for 46% of these top apps, while only 12.5% of the top smartphone apps fall into this category. The relevant TV apps also significantly outnumber the smartphone apps in the *News & Magazines* and *Sports* categories. We further crawled the descriptions of these apps from *Google Play Web*[2]. Figure 2 depicts the word cloud constructed from the descriptions of these apps. Several phrases (e.g., file, share, photo, friend, work, etc.) are conspicuous in Figure 2(a) but are discarded or less frequently appeared in Figure 2(b), which underlines the usage distinction between TV and telephone even more. On the other side, Google Play Store in smartphones lists 39 unique categories that are 27 categories more than the count in Google Play Store in TV devices. These experimental findings demonstrate that there is, in fact, a difference between the desired use scenarios for smartphone and smart TV apps.

[1] https://github.com/DannyGooo/AndroidTV-AndroidPhone-Collection

[2] https://play.google.com/store/apps

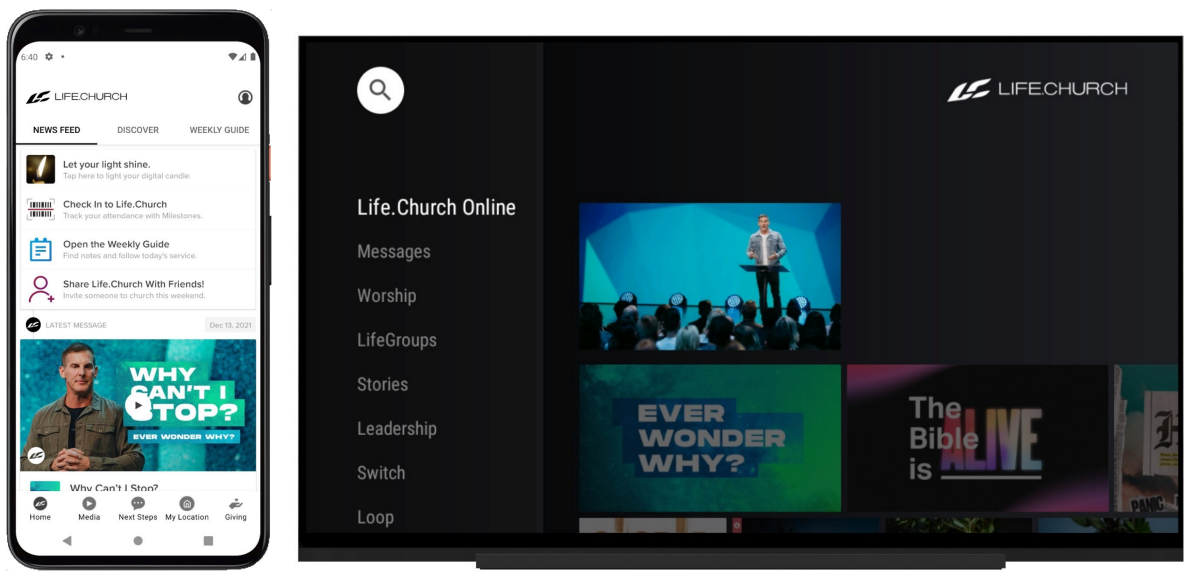

(a) Smartphone (b) Smart TV

**Figure 3:** app, Life.Church, with same SHA256 for base.apk installed on Smartphone and Smart TV.

Apart from the quantitative differences between smartphone and TV apps, smartphone apps exhibit qualitative differences in a variety of ways, as observed in the preliminary case study, despite the fact that both types of apps are developed in the same programming language (i.e., Java or Kotlin) and run on the same operating system (i.e., Android). In our daily life, TVs serve a distinct role from other devices, and they lack hardware features that other Android-powered devices generally offer. Some hardware features on other Android devices, such as touch displays, cameras, and GPS receivers, are not available on TVs. As introduced in Google Developer Doc [16], TVs are fully reliant on third-party hardware components, and users must utilize a remote control or a gaming pad to engage with TV apps.

As seen in Figure 3, the screenshots of the home page of the app named *"Life.Church"*, in phone and TV devices, respectively, are completely different. However, both the phone and TV versions share the same package name, *church.life.app*[3], and packaged together in one *App Package File (APK)*). Due to the hardware capabilities and screen size differences between smartphones and TVs, the appearance and functionality of the same app may vary across phone/TV devices. Additionally, there are apps in a pair with the same package name but distributed in different APK files (i.e., with different SHA256). The particular APK can only be distributed to the matching devices as the introduction of Android App Bundles (AAB)[4] on Google Play. For instance, *Facebook*[5] distribute different app on different platforms (i.e., TV and Phone). The phone's version of *Facebook* is known as an app for social with the size of 61.5 Mb. However, one live video streaming app with the size of 1.6 Mb is provided for the TV version of *Facebook* that is not collected by the well-known Android mobile app dataset, AndroZoo [3, 2]. Thus, the phone and TV versions of the same pair may be substantially different. As a result, security and privacy issues may occur differently depending on the device type.

The observations (including the usage scenario, the distinct apps with the same package name, and potential variances in security and privacy issues) drive us to examine further how smartphone and TV apps differ. We expect that developers and academics will be able to use our comparison study to properly design and analyze TV applications. Thus, we intend to delve deeper into the relationship between smartphone and TV apps by responding to the following research questions:

- **RQ1: [*Artefact*] To what extent are phone apps similar to their TV counterparts?** As we discussed above, the hardware on the TV is significantly different from that on Android mobile devices. This fact causes the user interaction paradigm for TV to be very different from that for smartphones. However, the similarities and differences in the applications between the two distinct devices have not yet been comprehensively examined. In this RQ, we analyze the non-code and code components of phone/TV pairs and want to find out to what extent phone apps are similar to their TV counterparts. This allows researchers and developers to better understand the relationships between smartphone apps and smart TV apps. It also serves as a preliminary but necessary step towards developing any smartphone-to-TV app migration strategy.

- **RQ2: [*Security and Privacy*] To what extent are the security/privacy issues different between phone/TV app pairs?** Android is now one of the primary targets of cyberattacks because of its popularity and openness [10]. Therefore, numerous prior studies have comprehensively investigated the security and privacy issues in Android applications. Nevertheless, the differences in security and privacy issues between Android mobile and TV applications are still poorly understood. In this RQ, we aim to determine to what extent the security and privacy risks differ between phone and TV applications. To achieve that, we compared the potential security vulnerabilities of Android TV and mobile applications from multiple perspectives.

- **RQ3: [*User Interaction*] How are user interactions varied between phone/TV app pairs?** Android is an event-driven system, and most of the methods are triggered by user interaction (e.g., press a button). However, the hardware of mobile phones and smart TVs is dissimilar, providing different interactions for app users. We propose this RQ to understand how user interactions differ between phone and TV apps. To answer this RQ, we analyze and compare the callback

[3] https://play.google.com/store/apps/details?id=church.life.app
[4] https://developer.android.com/guide/app-bundle
[5] com.facebook.katana

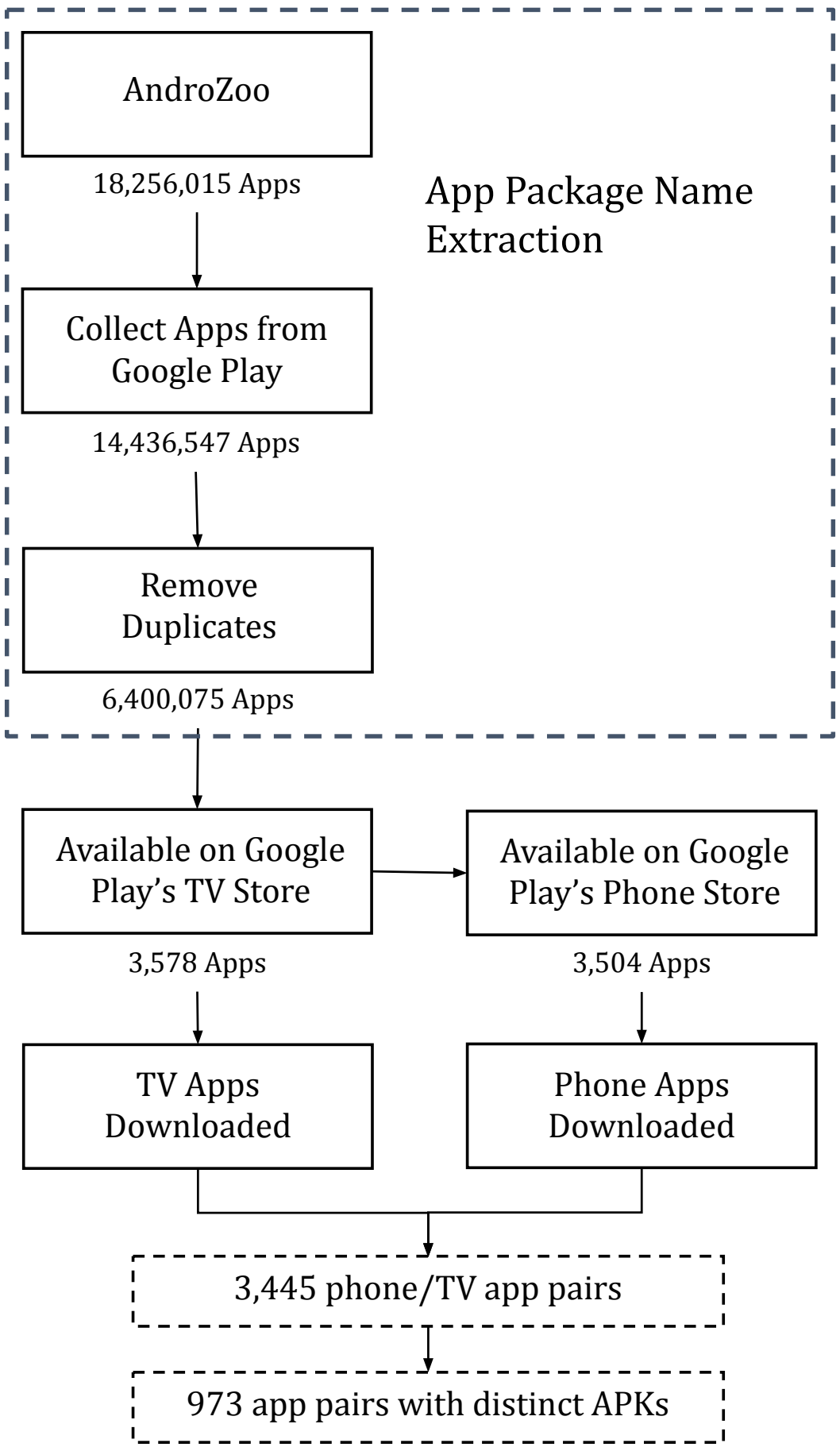

**Figure 4:** Working Process to Collect Android phone/TV app pairs from AndroZoo dataset.

event management approaches in the studied app pair dataset.

## 3. Dataset Collection and Preliminary Study

In this section, we present the process we have leveraged to collect real-world Android smartphone and smart TV app pairs (cf. Section 3.1). After that, we present the characteristics of the collected dataset through a preliminary study (cf. Section 3.2).

### 3.1. Dataset collection

To the best of our knowledge, there is no publicly accessible dataset of phone/TV app pairs; hence, as the first step of this work, we prepare a dataset containing phone/TV app pairs from scratch. The working process of collecting the app pairs contains the following two steps: 1) Retrieving the package names of the phone/TV apps; and 2) Downloading the phone and TV apps from the official Google Play store, respectively. Figure 4 illustrates the overall approach with the number of apps obtained in each step for quick reference. Details of each step are explained as follows.

**App Package Name Collection.** The official Google Play store does not indicate on the listing page which platforms (e.g., phone, TV, watch, etc.) an app works on. Therefore, we cannot know if an app has a TV version by the information provided on Google Play web pages. Our idea for collecting as many phone/TV app pairs as possible is to firstly get the list of apps available on the Google Play store and check if these apps are available both on smartphones and smart TVs. However, it is non-trivial to exhaust the Google Play store to get a complete list of apps. To work around this challenge, instead of crawling the whole Google Play store, we obtain the list of apps from AndroZoo [3, 2]. AndroZoo is an Android app repository that contains more than 18 million Android smartphone apps (and is still growing) collected from various app markets, including the official Google Play store.

We filter out the apps from alternative app stores and only keep the ones from the official Google Play store (approx. 14.5 million). After further removing the duplicated versions (i.e., historical versions of the same app), we obtain 6,400,075 unique package names.

We then check if the collected package names are available both on smartphones and smart TVs. According to the official Android documentation [37], the app developers may choose to exclude their apps from the Google Play store on specific devices (e.g., TV, watch, car, etc.) and hence will not be displayed or installed on these devices. Our idea is to simulate the desired devices (i.e., smart TV and smartphone, both with Android API level 30 and x86-based architecture) when querying the Google Play store. We also exclude the paid apps and only keep the free ones. Finally, we gathered 3,445 pairs of apps that have both free smartphone and smart TV versions.

It is worth mentioning that our approach can only collect app pairs that have identical package names on both platforms. While it is possible that a developer may name the same app on different platforms differently, it is non-trivial to determine if two apps with different package names are just the same app optimized for different platforms. Therefore, our dataset may miss app pairs that have different package names on smartphones and smart TVs. Nevertheless, to give readers an implication of how many app pairs our approach may miss, we further conducted an experiment to show the proportion of phone/TV app pairs that have the same package name. To this end, we retrieve the package names of all TV apps on an alternative Android app store APKMirror[6], which keeps the information of an app's compatible platforms (e.g., TV, watch, car, etc.). We have collected in total 819 unique TV apps from APKMirror, within which 667 apps are available on Google Play. Within the 667 TV apps, approx. 75% (i.e., 502) of them have a smartphone version with an identical package name. We further confirmed manually (by searching the other apps from the same developers, comparing the descriptions and running the apps on the TV and phone emulators) that approx. 11% (i.e., 72) of them have a smartphone version with different package names, while approx. 14% (i.e., 93) of them do not have a smartphone version. The results suggest that only a small percentage of apps (i.e., 11%) use different package names on different platforms that we may miss in our dataset.

[6] https://www.apkmirror.com/

**Table 1**
The metadata for app pairs based on category distribution

| Category | #. of Apps | #. of Apps With Ads | #. of Apps With IAPs | #. of Ratings | Rating (Star/5) | #. of Downloads |
|---|---|---|---|---|---|---|
| Entertainment | 1,059 | 249 (23.5%) | 222 (21.0%) | 33,342 | 3.92 | 6,207,614 |
| Music & Audio | 510 | 434 (85.1%) | 44 (8.6%) | 13,805 | 4.45 | 867,374 |
| Game | 364 | 260 (71.4%) | 184 (50.5%) | 80,609 | 3.93 | 4,653,291 |
| Video Players & Editors | 331 | 60 (18.1%) | 44 (13.3%) | 49,121 | 3.95 | 24,548,497 |
| Tools | 244 | 64 (26.2%) | 83 (34.0%) | 280,184 | 3.95 | 181,611,404 |
| Health & Fitness | 148 | 14 (9.5%) | 88 (59.5%) | 5,040 | 4.18 | 233,090 |
| Lifestyle | 112 | 23 (20.5%) | 58 (51.8%) | 11,303 | 4.36 | 478,347 |
| Sports | 108 | 70 (64.8%) | 39 (36.1%) | 34,211 | 4.08 | 2,986,739 |
| News & Magazines | 105 | 73 (69.5%) | 16 (15.2%) | 12,774 | 4.14 | 1,012,987 |
| Education | 102 | 40 (39.2%) | 65 (63.7%) | 749 | 4.21 | 138,980 |
| Business | 72 | 8 (11.1%) | 5 (6.9%) | 17,846 | 4.05 | 3,092,281 |
| Productivity | 63 | 8 (12.7%) | 12 (19.0%) | 5,835 | 4.17 | 1,148,480 |
| Communication | 50 | 4 (8.0%) | 11 (22.0%) | 202,045 | 4.21 | 127,799,849 |
| Food & Drink | 36 | 24 (66.7%) | 22 (61.1%) | 21,956 | 4.26 | 1,440,536 |
| Personalization | 24 | 7 (29.2%) | 5 (20.8%) | 3,607 | 3.88 | 731,861 |
| Books & Reference | 23 | 10 (43.5%) | 14 (60.9%) | 13,018 | 4.35 | 770,590 |
| House & Home | 13 | 0 (0.0%) | 2 (15.4%) | 203 | 3.91 | 58,314 |
| Art & Design | 12 | 3 (25.0%) | 1 (8.3%) | 240 | 3.93 | 34,007 |
| Travel & Local | 11 | 8 (72.7%) | 4 (36.4%) | 304 | 3.99 | 65,557 |
| Photography | 10 | 2 (20.0%) | 7 (70.0%) | 4,776 | 4.23 | 398,506 |
| Finance | 10 | 6 (60.0%) | 5 (50.0%) | 7,716 | 4.29 | 578,228 |
| Other | 38 | 17 (44.7%) | 11 (28.9%) | 3,532,616 | 4.34 | 196,937,355 |
| **Total** | 3,445 | 1,384 (40.2%) | 942 (27.3%) | 92,918 | 4.07 | 22,045,675 |

**Phone/TV App Pairs Collection.** After collecting the package names, we resort to the same methods used in the package name gathering procedure to query Google Play Store from the desired devices. We leverage an open-sourced project[7] for querying the Google Play Store. However, only limited devices are provided in this open-sourced project for querying. To add new devices options, we write scripts to extract particular device settings (e.g., Build.DEVICE, Build.MODEL, etc.) as parameters in the querying process. We simulate the same devices (i.e., phone/TV devices with API 30 and x86-based architecture) as the package name preparation process. To ensure version consistency, we write a script to download phone/TV apps in each pair simultaneously. We make our datasets and scripts publicly available[8] so that others can replicate our work. We also collect the metadata and user reviews for further characterizing the dataset.

## 3.2. Data Characteristics

We now present a preliminary study to show the metadata and user review of the collected 3,445 phone/TV app pairs. The metadata and user review are crawled from the Google Play Store web. We note to the readers that the values for some metadata (e.g., *Size*, *Current Version*, etc.) in some apps' Google Play Store pages might not be unique and hence are marked as "Varies with device". In such a case, we have to exclude those cases when characterizing the metadata values of the collected apps. However, if an actual value is assigned to given metadata, we will consider the value is the same across all possible device versions. In this section, we describe the characteristics of our dataset in terms of (1) Metadata and User Review, and (2) Pairs with different apps.

[7]https://github.com/NoMore201/googleplay-api
[8]https://github.com/DannyGooo/AndroidTV-AndroidPhone-Collection

**Metadata and User Review.** Table 1 shows the metadata present in Google Play of the collected app pairs grouped by category. The second column of the table shows the number of apps that fall into the corresponding categories, followed by the third and fourth columns presenting the number of apps featuring advertisement and in-app purchase (IAP[9]). Advertisements and IAP are the main revenue streams for phone and TV apps. In general, around 40% of apps have in-app advertising. Apps in specific categories, such as Music & Audio, Travel & Local, and Games exhibit a higher chance of including advertisements (i.e., 85.1%, 72.7%, and 71.4%, respectively). On the other hand, several app categories have a lower chance of presenting in-app advertising. For instance, all apps collected from House & Home category are ad-free, and only 8.0% of Communication apps contain advertisements. There are around 27% of apps feature IAP. IAPs are more likely to appear in Photography and Education apps (i.e., 70% and 63.7%, respectively). It is interesting to observe that apps with advertisements are more likely to have IAPs. Particularly, 34.2% of apps with advertisements also have IAPs, while only 22.7% of apps without advertisements contain IAPs. In addition, columns 5, 6 and 7 present the average number of user reviews in each category, the average star rating (out of 5) of each app, and the average number of downloads, which shows the popularity of the apps to some extent. Note that the excessive number of ratings and downloads in the *Other* category and *Facebook* app that falls in this category.. The main reason is that the *Facebook* app performs distinct functionalities on TV and mobile phones, and Google Play's TV app store doesn't include a "Social" category.

**App pairs with identical/different APKs.** We compare the SHA256 of the apps in each pair and interestingly find that a large proportion of pairs deploy the same APK for TV and phone versions. As demonstrated in Figure 5, there are total $2,472$ pairs that have identical APKs on phones and TVs, which implies that the code and resources in each phone/TV version are merged into one single APK file. In contrast, 973 pairs were deployed with the different APKs. In general, more developers tend to package both the phone and TV version in a single APK, while for apps in particular categories (e.g., Music & Audio, Sports, News & Magazines, etc.), more developers choose to develop individual APKs for different platforms.

Additionally, we further investigated the size of the APKs in phone/TV app pairs. The average phone and TV app sizes for the collected apps are 27.2 and 25.9 Mb, respectively. As shown in Figure 6, for the apps with a single APK for both platforms, the average APK size is 31.1 Mb. On the other hand, the size of the APKs with different APKs on the phone and TV is significantly smaller (i.e., 12.6 Mb and 17.1 Mb for TV and phone apps, respectively). Interestingly, the size (i.e., 31.1 Mb) for a single APK on both platforms is comparable with the sum of the APK size (i.e., 29.7 Mb) for the phone/TV versions with different APKs, which further suggests that apps that have single APKs on TVs and phones

[9]https://support.google.com/googleplay/answer/1061913

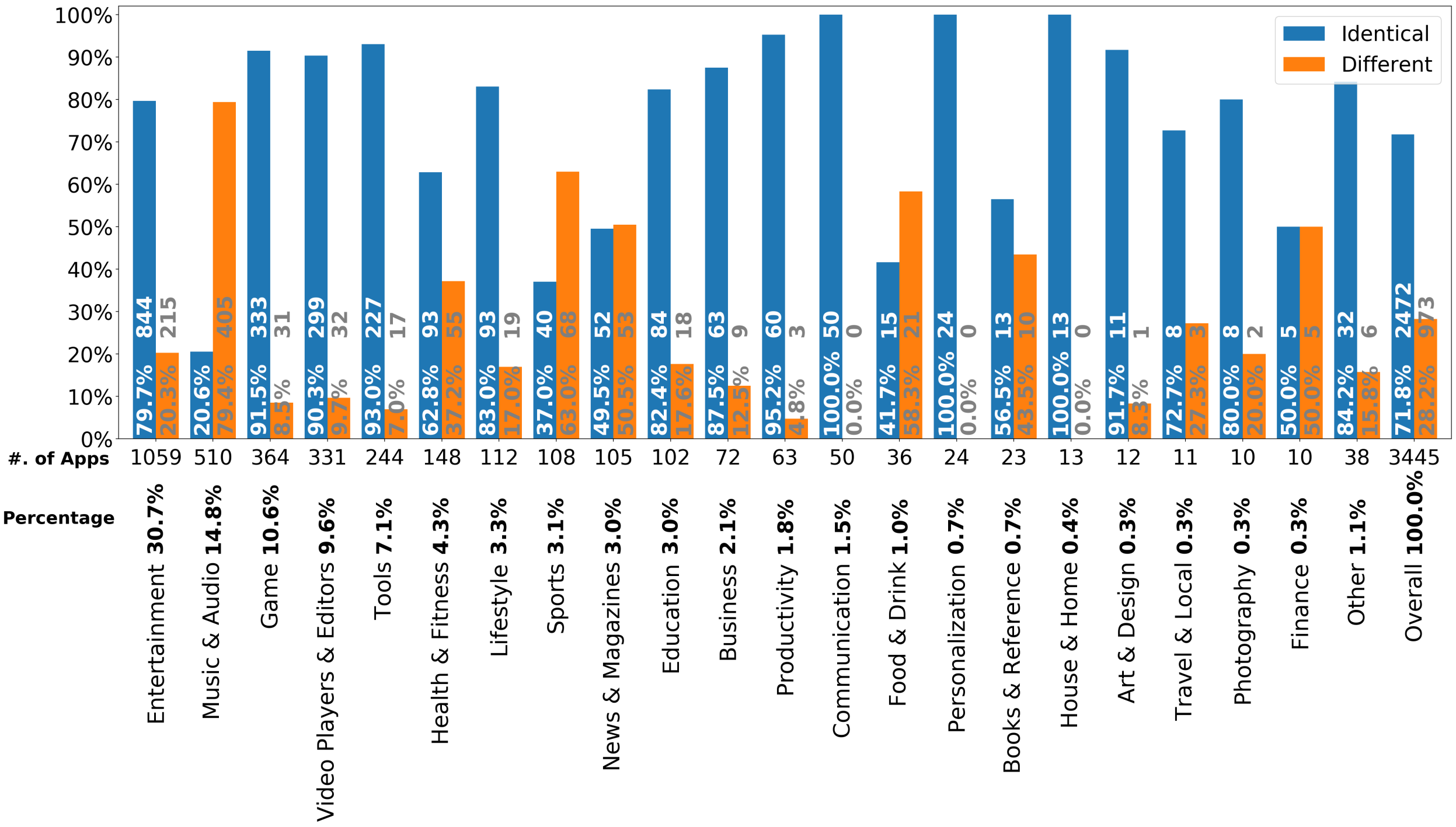

**Figure 5:** Category Distribution of pairs with Different and Identical APK

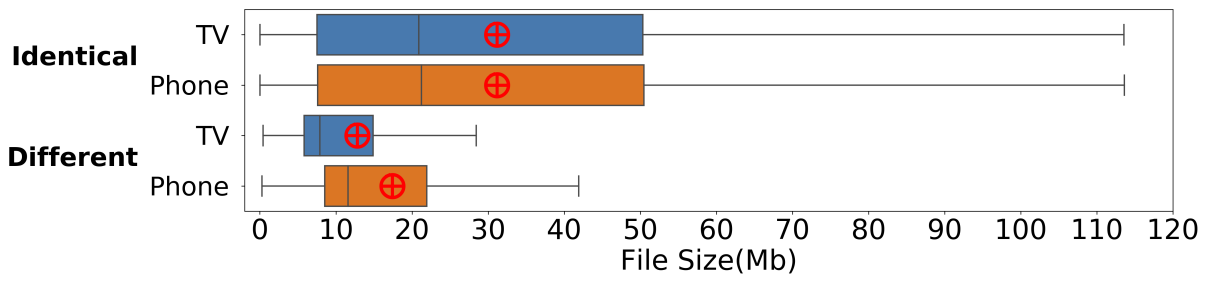

**Figure 6:** Distribution of App size

tend to package code and resource on both platforms into one APK.

## 4. Empirical Results

Note that our empirical analyses focus on comparing the phone/TV app pairs from three perspectives (i.e., APK artefact, security and privacy, and user interaction). We disregard the apps whose TV and phone versions use the same apk (i.e., 2,472 identical phone/TV pairs) since prior studies [31, 41] have investigated the single phone/TV-enabled APK. Thus our study compares the app pairs that have a different apk for each version. According to our results in Section 3.2, there are 973 pairs with different apps. The sub-sections of this section report our study's findings for the proposed three research questions.

### 4.1. RQ1: APK Artefact

In this RQ, we examine the similarity of the non-code (e.g., resources) and the code (e.g., Activities, Permissions) components of phone/TV app pairs. Resources are non-code assets that the app code could access, such as images, textual data, and User Interface (UI) layouts. In terms of the app code, both components and methods are inspected. Android components are the essential building blocks of an Android app, and each component has a well-defined lifecycle. An Android app has four primary components, including Activities, Services, Broadcast Receivers, and Content Providers. An Activity is a User Interface with a back-end class that deals with the action that is done on the User Interface. A Service is a back-end class that does not have a User Interface. The Broadcast Receiver is primarily used to handle broadcast intent from the Android operating system or other apps. A Content Provider is a location where an app's data may be stored and accessible by other apps. Methods are blocks of code that have instructions for doing certain things that are run when the method is called. Apart from the components and the methods, we also compare the permissions declared in each phone/TV app pair. Permissions must be declared in the manifest file before accessing sensitive user data (e.g., location and contact list), as well as certain system features (e.g., camera and Internet). Comparing the permissions can help us understand the feature between app pairs.

**Similarity Analysis.** We leverage SimiDroid [23] to compare the pairwise similarities and differences between apps in each phone/TV pair. SimiDroid is a state-of-the-art approach to identify and explain similarities in Android applications. SimiDroid initially parses the app pairs and organizes them into key/value mappings. Depending on the comparison level (i.e., resource, component, or method), various attributes are extracted to create their keys and values. After extracting the key/value mapping pairs (map1 and map2), the following metrics are used to compare them: (1) identical, when compared key/value pairs are exactly matched; (2) similar, where the key is the same but values differ; (3) new, where the key exists only in map2 (i.e., the block only exists in the TV version); and (4) deleted, where the key only exists in map1(i.e., the block only exists in the smartphone version). Finally, the similarity score is calculated as:

$$similarity = max\{\frac{identical}{total-new}, \frac{identical}{total-deleted}\}$$

where:

$$total = identical + similar + new + deleted$$

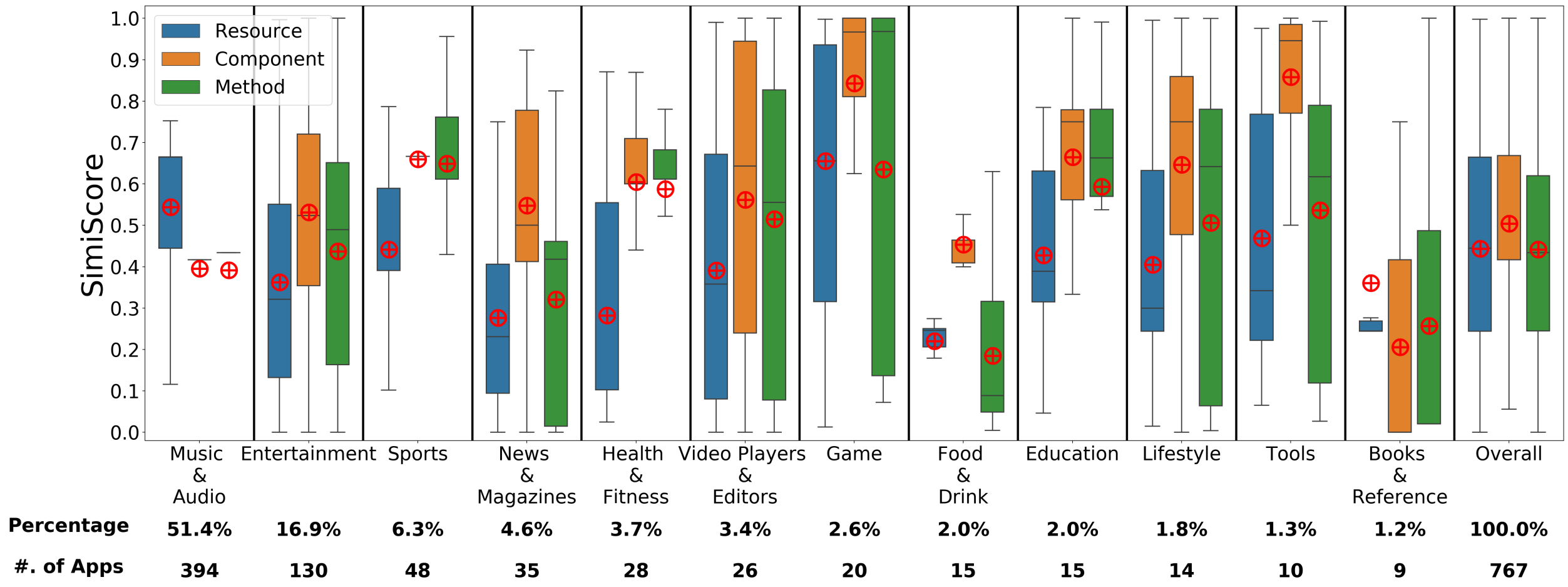

**Figure 7:** Distribution of SimiDroid Score for Pairs with Different APKs

The similarity scores for the phone/TV pairs are shown in Figure 7 from a variety of perspectives, including resource, component, and method based on the app's category. On average, about 44.6% of resource files, 50.4% of component blocks, and 44.1% of method blocks are reused while designing multi-platform apps for smartphone and TV devices. Further examination of the file types reveals that the most often reused file types are PNG (e.g., UI widgets) and XML (e.g., UI layouts and string values), accounting for 70.8% and 11.8%, respectively, of all reused files. In general, the resources are the most altered part in the apps throughout most categories (excluding Music & Audio and Books & Reference), compared with the code (i.e., components and methods). The high similarity in code indicates that the essential functionality between the apps of each pair is comparable. This observation implies that despite the hardware and functionality differences, a certain amount of code is migrated from the smartphone to the TV version without modification.

**Permission Analysis.** The app permission framework is a critical component of the Android ecosystem since it ensures the security of Android users' privacy. An Android app must seek the user's permission when accessing sensitive user data (e.g., location and audio). The permissions have to be explicitly declared in the *AndroidManifest.xml* file of an Android app. According to the Android Developer Guide, permissions can be classified as *normal*, *signature*, or *dangerous* according to their riskiness [13]. *Normal* permission is lower-risk permission that would be granted at installation without approval. It provides requesting apps with access to isolated application-level features with minimal risk to other apps, the system, or the user. *Dangerous* permissions are higher-risk permissions that would give the requesting app access to private user data or control over the device, which can negatively impact the user. Due to the inherent danger associated with this type of permission, the system does not automatically grant it to the requesting app. Instead, such permissions need to be granted at run-time. *Signature* permission will only be automatically granted if the requesting app is signed with the same certificate as the app that declared the permission. In this study, we compare the differences in the declared permissions for each phone/TV app pair to understand the kind of functionalities and resources the app may access. To this end, we utilize the Android Asset Packaging Tool[10] (AAPT), a well-known android static analysis tool, to extract the declared permissions in the app pairs.

Table 2 lists the most frequently declared permissions in phone and TV apps. Overall, TV apps declare significantly fewer permissions than phone versions. On average, TV apps declare nearly half the number of permissions as their phone counterparts (i.e., each phone app declares 13.9 permissions while the corresponding TV version declares 7.3 permissions). The number of apps requesting network-related permissions (e.g., *INTERNET* and *ACCESS_NETWORK_STATE*) in phone and TV apps are comparable. It does make sense as apps on both devices require Internet access. The number of apps containing other permissions, on the other hand, significantly decreases in the TV versions. For example, 550 mobile phone apps declare the permission *ACCESS_COARSE_LOCATION*, while only 106 TV apps include this permission. This observation indicates that many features are removed from the TV versions, such as pushing notifications, accessing locations, recording audio, etc. A possible explanation for this result might be the lack of hardware features on TVs (e.g., GPS, microphone). As mentioned in [16], some of the hardware is usually not available on TV devices (e.g., camera, microphone, touch screen, etc.), and the developers should avoid using the related features in TV apps. However, the associated permissions are still found to be declared in some TV apps, which may result in hindering the user experience. This finding suggests that tools are needed before an app is released on the Google Play Store to avoid these kinds of issues.

[10] https://developer.android.com/studio/command-line/aapt2

**Table 2**
Top 15 permissions in phone and TV apps for Pairs with Different APKs

| Permissions in **TV** Apps | Count | Protection Level | Permissions in **Phone** Apps | Count | Protection Level |
|---|---|---|---|---|---|
| INTERNET | 952 (92.0%) | Normal | INTERNET | 962 (92.9%) | Normal |
| ACCESS_NETWORK_STATE | 941 (90.9%) | Normal | ACCESS_NETWORK_STATE | 961 (92.9%) | Normal |
| WAKE_LOCK | 781 (75.5%) | Normal | WAKE_LOCK | 941 (90.9%) | Normal |
| FOREGROUND_SERVICE | 517 (50.0%) | Normal | RECEIVE[2] | 874 (84.4%) | Signature |
| BIND_GET_INSTALL_REFERRER_SERVICE[1] | 383 (37.0%) | Normal | FOREGROUND_SERVICE | 833 (80.5%) | Normal |
| ACCESS_WIFI_STATE | 312 (30.1%) | Normal | VIBRATE | 586 (56.6%) | Normal |
| RECEIVE_BOOT_COMPLETED | 274 (26.5%) | Normal | **ACCESS_COARSE_LOCATION** | 550 (53.1%) | **Dangerous** |
| RECEIVE[2] | 255 (24.6%) | Signature | **ACCESS_FINE_LOCATION** | 544 (52.6%) | **Dangerous** |
| BILLING[3] | 232 (22.4%) | Normal | GET_ACCOUNTS | 514 (49.7%) | Normal |
| **READ_EXTERNAL_STORAGE** | 189 (18.3%) | **Dangerous** | **READ_PHONE_STATE** | 495 (47.8%) | **Dangerous** |
| **WRITE_EXTERNAL_STORAGE** | 175 (16.9%) | **Dangerous** | BIND_GET_INSTALL_REFERRER_SERVICE[1] | 488 (47.1%) | Normal |
| **RECORD_AUDIO** | 117 (11.3%) | **Dangerous** | **WRITE_EXTERNAL_STORAGE** | 451 (43.6%) | **Dangerous** |
| **ACCESS_FINE_LOCATION** | 116 (11.2%) | **Dangerous** | **RECORD_AUDIO** | 433 (41.8%) | **Dangerous** |
| **ACCESS_COARSE_LOCATION** | 106 (10.2%) | **Dangerous** | ACCESS_WIFI_STATE | 422 (40.8%) | Normal |
| CHANGE_WIFI_STATE | 102 (9.9%) | Normal | RECEIVE_BOOT_COMPLETED | 401 (38.7%) | Normal |

[1] *com.google.android.finsky.permission.*BIND_GET_INSTALL_REFERRER_SERVICE
[2] *com.google.android.c2dm.permission.*RECEIVE
[3] *com.android.vending.*BILLING
All other pemissions are defined by Android, starting with *android.permission.*

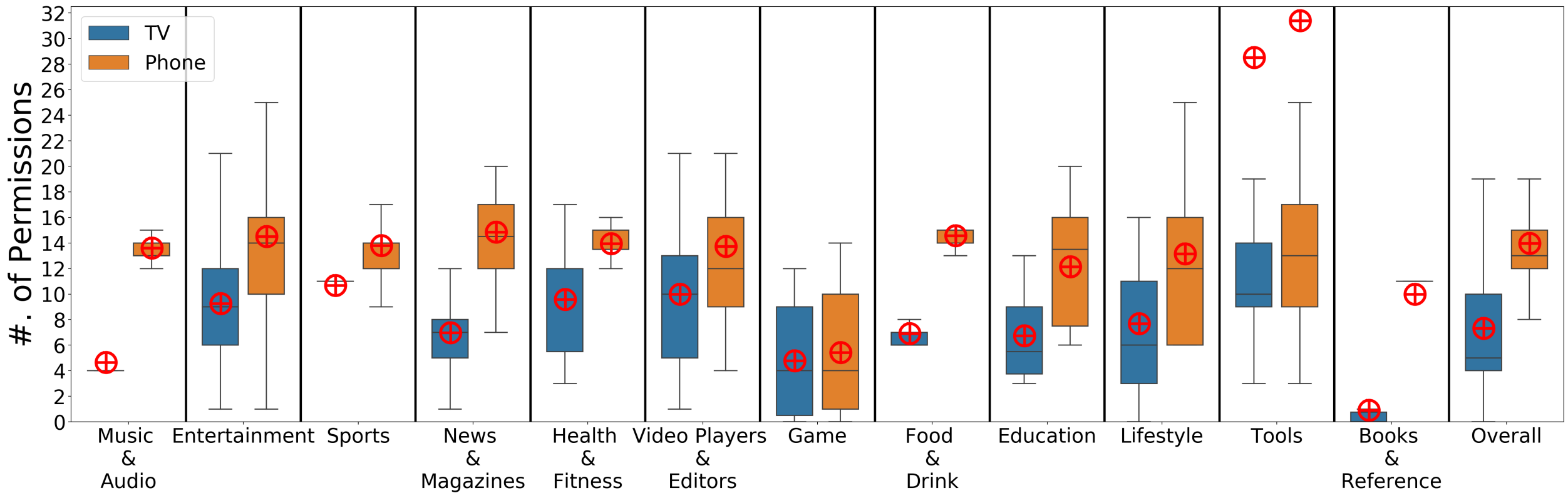

**Figure 8:** Distribution of the count of the type of permissions in phone and TV apps for Pairs with Different APKs

Let us take an in-depth look at the permissions declared in each app category. Figure 8 shows a category-wise breakdown of the count of permissions in each app. It is not surprising to observe that the average number of permissions declared in each category of TV apps is less than its phone counterparts due to the lack of hardware features in TV devices. Interestingly, in some categories (e.g., Game and Tools), the number of permissions declared in the TV apps are close to that in their phone versions, while some (e.g., Music & Audio and Books & References) show dramatic differences. This observation demonstrates that TV/phone app pairs in specific categories may demonstrate similar functionalities and features, while in some categories, many features are removed from the TV versions. We perform a further detailed analysis on the permissions declared in the Music & Audio category. The top four permissions of TV apps in this category are declared by at least 80% of these apps, and the remaining permissions are all declared in less than 10% of these apps. These top four permissions are exactly the same as the top four TV permissions listed in Table 2. In comparison, there are 11 permissions declared by at least 88% of smartphone apps in this category. The permissions that have been most frequently removed from the smart TV apps in Music & Audio category are registering and receiving messages from Google Cloud Messaging, reading phone states, vibrating, recording audio, accessing location, and accessing accounts. These findings are also consistent with previous reports that hardware-intensive features (e.g., vibrating, obtaining location information, etc.) are often removed from TV apps.

**Table 3**
Duplicated Permissions in phone and TV apps for Pairs with Different APKs

| Duplicated Permissions in **TV** Apps | **Count** | Protection Level | Duplicated Permissions in **Phone** Apps | **Count** | Protection Level |
|---|---|---|---|---|---|
| ACCESS_NETWORK_STATE | 30 | Normal | WAKE_LOCK | 349 | Normal |
| INTERNET | 29 | Normal | INTERNET | 12 | Normal |
| **READ_PHONE_STATE** | 24 | **Dangerous** | ACCESS_NETWORK_STATE | 7 | Normal |
| WAKE_LOCK | 21 | Normal | **READ_PHONE_STATE** | 3 | **Dangerous** |
| **WRITE_EXTERNAL_STORAGE** | 2 | **Dangerous** | VIBRATE | 3 | Normal |
| | | | **WRITE_EXTERNAL_STORAGE** | 2 | **Dangerous** |
| | | | SYSTEM_ALERT_WINDOW | 1 | Signature |
| | | | **ACCESS_FINE_LOCATION** | 1 | **Dangerous** |
| | | | BILLING[1] | 1 | Normal |
| | | | FOREGROUND_SERVICE | 1 | Normal |
| | | | BLUETOOTH | 1 | Normal |

[1] *com.android.vending.*BILLING
All other pemissions are defined by Android, starting with *android.permission.*

**Declaration of Duplicated Permissions.** We then investigate how the TV/phone app pairs were implicated in the duplicated permission issues (i.e., the same permission is declared more than once in the manifest). According to our analysis, 366 smartphone apps and 31 TV apps declare the same permissions multiple times in their manifest file. We observed that certain apps may involve more than one type of duplicated permission. More apps are found with permission duplication in their phone versions than their TV counterparts. In contrast, on average, for apps with permission duplication, TV versions contain more duplication permission, with 3.4 but just one duplication permission contained in TV and phone versions, respectively. We next examined these permissions in further detail. As demonstrated in Table 3, the types of duplicated permissions are quite different between phone/TV app pairs. Particularly in the Music & Audio category, the *WAKE_LOCK* permission has been largely re-declared among smartphone apps, accounting for 84.3% of smartphone apps in this category, while only 3.7% of TV apps in this category are affected by such permission duplication. Most of the TV apps with duplicated permissions (22 apps, 66.7%) are developed by the same developer, *iNmyStream*, spanning three categories, including Entertainment, Music & Audio, and News & Magazines. In the phone apps, the package name of more than 90% of duplicated-permission smartphone apps (335 apps, 91%) begins with "com.icreo.", but are built by different developers. Duplicated permissions mainly indicate developers' poor development practices. Other than that, there is also a chance for app repackaging. As revealed by [25, 24], smartphone apps may have duplicated permission declarations. This kind of recurrence may be an indicative of app repackaging, since opportunists typically attach required permissions without verifying whether they have previously been declared or not, especially for the high occurrence of duplicated permissions in the TV apps. The above evidence suggests the necessity for developers to improve their development practices.

In summary, these results indicate that TV apps are less complicated than their smartphone counterparts, yet phone/TV app pairs have a high degree of similarity when comparing non-code and code components. We find that 43% of resource files and up to 50% of methods are reused between smartphone/TV app pairs. Our results imply that a certain amount of the source code of TV apps is directly duplicated from their smartphone counterparts. However, we discover that due to the platform differences, there are several obvious differences between phone/TV app pairs. The number of permissions that TV apps declare is almost half that of their phone counterparts. Compared to TV apps, there are more phone apps involving permission duplication.

**RQ 1: Summary**

- 43% of resource files and up to 50% of methods are reused between smartphone/TV app pairs.
- On average, TV apps declare nearly half the number of permissions as their phone counterparts.
- There are more phone apps involving permission duplication.
- On average, each TV app declares 3.4 duplicated permission, while this number in phone apps is one.

## 4.2. RQ2: Security & Privacy Analysis

In this section, we analyze the security and privacy risks in phone/TV pairs. We resort to AndroBugs [27] to understand the potential security vulnerabilities. To study privacy breaches (i.e., data leaks) in Android TV apps, we examine these 973 phone/TV pairs using FlowDroid [5].

**AndroBugs Analysis.** AndroBugs reports the details of vulnerabilities discovered in each app, categorizing them as *Critical*, *Warning*, *Notice*, or *Info*, according to the harm they may bring. Overall, AndroBugs reports that each app in phone/TV app pairs has at least 52 vulnerabilities. Figure 9 demonstrates the number of critical vulnerabilities detected in the collected phone/TV pairs. Among all the 973 phone/TV pairs, 961 phone apps, and 935 TV apps are flagged to contain critical security vulnerabilities. Over half of the apps contain at least three critical issues, and the average number of critical issues is 3 and 2.5 for phone and TV apps, respectively. In general, the phone version in the app pair contains more critical vulnerabilities, except for approx. 11% of TV apps in the pairs contain more critical vulnerabilities than their phone counterparts. The majority of these apps are from Music & Audio, Entertainment, and Food & Drink categories. This evidence indicates that the majority of Android TV apps have fewer security vulnerabilities than phone apps, but in some specific categories, TV apps would suffer more vulnerabilities than their phone counterparts. It is interesting to observe that more than 60% of TV apps in the Food & Drink category contain more critical vulnerabilities than phone apps. Considering the fact that TV apps contain fewer Activities and code, the significance of critical issues in Android TV apps should not be overlooked.

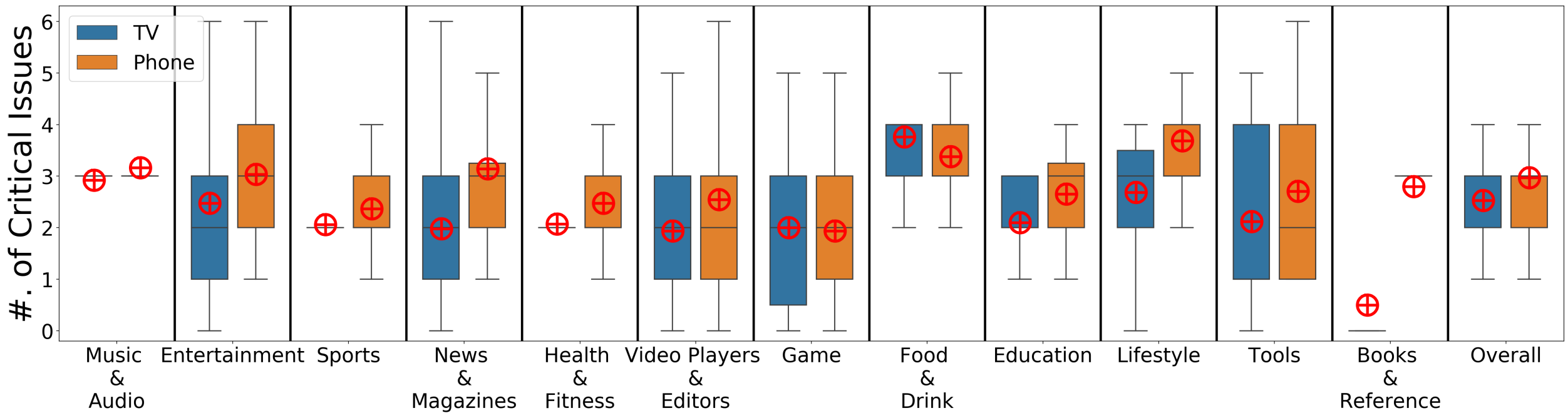

**Figure 9:** Category Distribution of AndroBugs Critical Issue in phone and TV apps for Pairs with Different APKs

**Table 4**
Distribution of Critical Vulnerabilities Found by AndroBugs for Pairs with Different APKs

| Type | Detailed Vulnerability Descriptions | Music & Audio | Entertainment | Sports | Health & Fitness | News & Magazines | Food & Drink | Others | **Total** |
|---|---|---|---|---|---|---|---|---|---|
| | | | | | **TV** | | | | |
| SSL_Security | SSL Connection | 403 | 203 | 67 | 53 | 50 | 21 | 120 | 917 |
| | Verifying Host Name in Custom Classes | 344 | 9 | 1 | 0 | 2 | 15 | 9 | 380 |
| | SSL Certificate Verification | 20 | 23 | 0 | 0 | 2 | 2 | 4 | 51 |
| Implicit_Intent | Implicit Service | 337 | 43 | 10 | 30 | 7 | 2 | 30 | 459 |
| WebView | WebView RCE Vulnerability | 14 | 102 | 11 | 20 | 22 | 4 | 72 | 245 |
| AndroidManifest | ContentProvider Exported | 6 | 72 | 7 | 4 | 9 | 3 | 31 | 132 |
| Command | Runtime Command | 48 | 22 | 0 | 1 | 4 | 1 | 9 | 85 |
| Permission | App Sandbox Permission | 6 | 6 | 39 | 0 | 4 | 1 | 9 | 65 |
| KeyStore | KeyStore Protection | 0 | 25 | 1 | 2 | 2 | 13 | 10 | 53 |
| Encryption | Base64 String Encryption | 3 | 5 | 2 | 0 | 1 | 14 | 7 | 32 |
| Fragment | Fragment Vulnerability | 2 | 10 | 2 | 3 | 1 | 2 | 9 | 29 |
| | Others | 1 | 12 | 0 | 1 | 0 | 1 | 4 | 19 |
| | **Total** | 1184 | 532 | 140 | 114 | 104 | 79 | 314 | 2467 |
| | | | | | **Phone** | | | | |
| SSL_Security | SSL Connection | 385 | 207 | 68 | 54 | 53 | 21 | 128 | 916 |
| | Verifying Host Name in Custom Classes | 53 | 12 | 1 | 1 | 2 | 0 | 15 | 84 |
| | SSL Certificate Verification | 4 | 15 | 1 | 1 | 2 | 2 | 13 | 38 |
| WebView | WebView RCE Vulnerability | 393 | 146 | 17 | 22 | 49 | 19 | 100 | 746 |
| Implicit_Intent | Implicit Service | 390 | 122 | 20 | 47 | 24 | 18 | 82 | 703 |
| Fragment | Fragment Vulnerability | 2 | 41 | 5 | 6 | 7 | 1 | 19 | 81 |
| AndroidManifest | ContentProvider Exported | 14 | 26 | 5 | 0 | 6 | 1 | 20 | 72 |
| Permission | App Sandbox Permission | 16 | 6 | 39 | 0 | 3 | 0 | 5 | 69 |
| Command | Runtime Command | 17 | 17 | 1 | 2 | 3 | 0 | 11 | 51 |
| KeyStore | KeyStore Protection | 0 | 24 | 2 | 2 | 3 | 2 | 12 | 45 |
| Hacker | Base64 String Encryptio | 5 | 9 | 1 | 0 | 12 | 3 | 12 | 42 |
| | Others | 1 | 14 | 1 | 1 | 1 | 3 | 9 | 30 |
| | **Total** | 1280 | 639 | 161 | 136 | 165 | 70 | 426 | 2877 |

We are continuing our investigation into these critical vulnerabilities. Table 4 outlines the various categories of critical vulnerabilities and their detailed descriptions in descending order of prevalence. Vulnerability types with fewer occurrences are classified as "Others". Now, we elaborate on three of the most common vulnerability categories in Android phone/TV app pairs:

- **SSL Security**: This kind of vulnerability accounts for $1,348$ ($55\%$, $917 + 380 + 51$) and $1,038$ ($36\%$, $916 + 84 + 38$) of all critical vulnerabilities found in the TV and phone apps, respectively. These flaws are inextricably linked to the apps' internet access mechanism. Malicious substances may be able to capture an app's information across the network if the app uses SSL incorrectly [14]. Android apps, for example, maybe vulnerable to *man-in-the-middle (MITM)* attacks if they connect to the internet without employing strong encryption (e.g., when using HTTP instead of HTTPS).

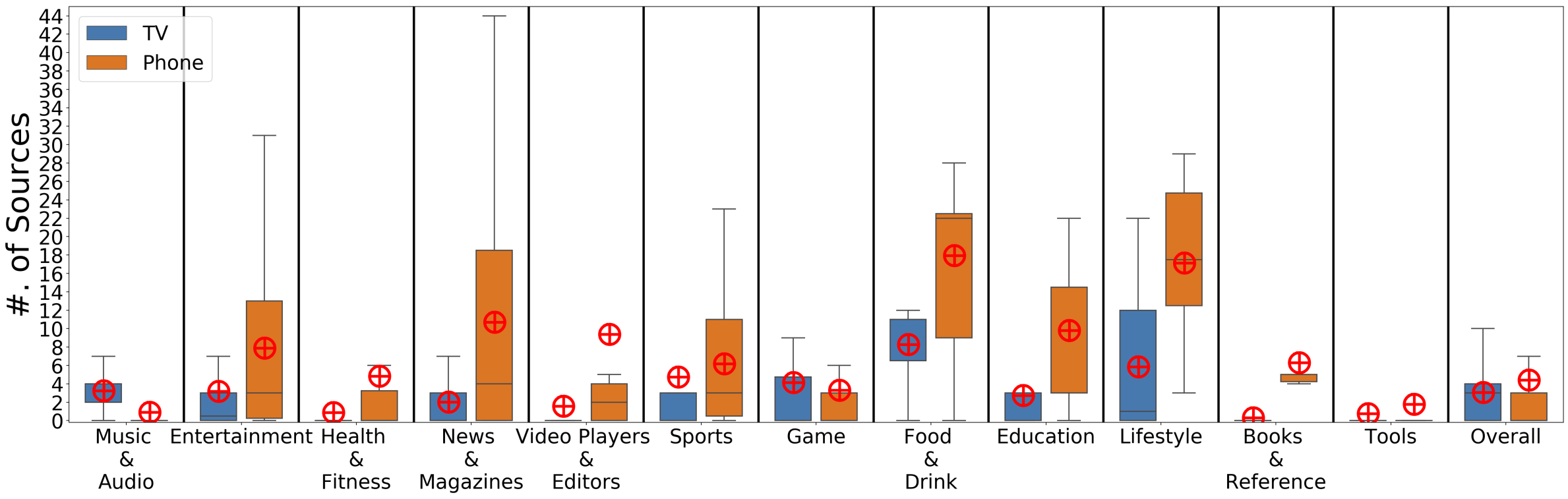

| | Music & Audio | Entertainment | Health & Fitness | News & Magazines | Video Players & Editors | Sports | Game | Food & Drink | Education | Lifestyle | Books & Reference | Tools | Overall |
|---|---|---|---|---|---|---|---|---|---|---|---|---|---|
| **Percentage** | **51.4%** | **16.9%** | **6.3%** | **4.6%** | **3.7%** | **3.4%** | **2.6%** | **2.0%** | **2.0%** | **1.8%** | **1.3%** | **1.2%** | **100.0%** |
| **#. of Apps** | **394** | **130** | **48** | **35** | **28** | **26** | **20** | **15** | **15** | **14** | **10** | **9** | **767** |

**Figure 10:** Distribution of FlowDroid Sources in phone and TV apps for 755 Pairs with Different APKs

- **WebView**: In total, there are 245 (10%) and 746 (26%) vulnerabilities in this type found in TV and phone apps, respectively. This is critical since an attacker may use this vulnerability to remotely manipulate the host app to run arbitrary Java code [12]. For example, attackers can execute Java code in the host apps, giving attackers access to command-line tools and posing additional security risks to users [40].

- **Implicit Intent**: This kind of vulnerability accounts for 459 (19%) and 703 (24%) of all vulnerabilities found. The Android ecosystem makes use of the Intent mechanism to facilitate the reuse of functionality [9]. However, implicit Intents may be intercepted by other components [35]. Therefore, it is a potential security concern since attackers may leverage implicit Intents to get access to sensitive information, posing threats to users' privacy.

As demonstrated in Table 4, SSL_Security is the most frequently detected critical vulnerability type in both the phone and TV apps. The second and the third most detected vulnerability types are also similar (i.e., *Implicit_Intent* and *WebView*, respectively, in TV apps, while *WebView* and *Implicit_Intent*, respectively in phone apps). Among the SSL_Security issues, the apps impacted by the vulnerability, *Verifying Host Name in Custom Classes*, would let attackers exploit a valid certificate to conduct MITM attacks without the users' awareness. As shown in Table 4, the TV versions in the app pairs are more likely to suffer from this vulnerability, with 84 and 380 vulnerabilities occurring in the phone and TV versions, respectively. This observation indicates that certain apps' TV versions are at significant security risk while the phone versions remain unaffected.

**FlowDroid Analysis.** To further spot privacy breaches in Android TV apps, we resort to FlowDroid [5] to examine the collected app pairs. FlowDroid identifies data flows from sensitive sources to potentially dangerous sinks, where sources are methods through which such data flows enter the app, and sinks are the methods from which they leave the app. Our results show that FlowDroid successfully finished scanning 785 phone apps and 910 TV apps. Approximately 39% (307 apps) of the 785 phone apps and 56% (512 apps) of the 910 TV apps contain at least one privacy leak, revealing that more TV apps are encountered privacy issues than their phone counterparts.

We continue our investigation of the sources revealed by FlowDroid in phone/TV pairs. The average number of sources detected in each phone app (4.3 sources) is greater than its TV version (3 sources), as illustrated in Figure 10. In general, phone apps have more sources than their TV counterparts, except for the pairs in the Music & Audio and Game categories. TV apps in Music & Audio category have an average of 3.2 sources, compared to only 0.9 sources in their phone counterpart. Table 5 lists the sources detected by FlowDroid in the phone/TV pairs. The sources detected with fewer than 20 occurrences are grouped in the *Others* category. *Android.content.pm.PackageManager.queryIntentServices* is the most leaked source (i.e., 1, 259 occurrence) among TV apps, contributing to more than half of the sources in the leaks (i.e., 2, 282 occurrence). Many apps utilize *queryInterntServices* to get all services that may match a certain intent and then wake these services up depending on the return values of *queryIntentServices*, which are solely used for inter-app communication [45]. The most leaked source in the phone apps is the *getString* method in the database package. This is unsurprising as the database often involves vast amounts of data, hence having more chance to leak the data. The geographical locations of the devices are also popular sources in the leaks. Although most of the TV devices do not have a GPS module, WiFi or network information can be used to approximate the device's location. In contrast to phone devices, the geographical locations of Android TVs are often fixed and expose users' home addresses. Attackers may utilize this information to get a wealth of additional information on users or even to pose physical threats [41, 44].

**Table 5**
Distribution of Sources Detected more than 20 times by FlowDroid for 755 Pairs with Different APKs

| **Sources** | Music & Audio | Entertainment | Food & Drink | Sports | Game | Lifestyle | Others | **Total** |
|---|---|---|---|---|---|---|---|---|
| | | | | TV | | | | |
| android.content.pm.PackageManager.queryIntentServices | 1197 | 13 | 0 | 14 | 0 | 13 | 22 | 1259 |
| android.database.Cursor.getString | 25 | 117 | 8 | 25 | 5 | 6 | 68 | 254 |
| android.location.Location.getLatitude | 0 | 84 | 0 | 29 | 22 | 24 | 35 | 194 |
| android.location.Location.getLongitude | 0 | 84 | 0 | 27 | 22 | 24 | 35 | 192 |
| java.util.Locale.getCountry | 5 | 2 | 115 | 0 | 9 | 0 | 2 | 133 |
| java.net.HttpURLConnection.getInputStream | 8 | 46 | 1 | 11 | 15 | 6 | 34 | 121 |
| java.net.URLConnection.getInputStream | 36 | 10 | 0 | 0 | 0 | 2 | 3 | 51 |
| android.content.pm.PackageManager.queryIntentActivities | 0 | 6 | 0 | 8 | 2 | 3 | 11 | 30 |
| android.content.pm.PackageManager.queryBroadcastReceivers | 0 | 5 | 0 | 7 | 2 | 4 | 5 | 23 |
| Others | 3 | 8 | 0 | 2 | 6 | 0 | 6 | 25 |
| **Total** | 1274 | 375 | 124 | 123 | 83 | 82 | 221 | 2282 |
| | | | | Phone | | | | |
| android.database.Cursor.getString | 166 | 234 | 4 | 31 | 4 | 17 | 409 | 865 |
| android.location.Location.getLongitude | 40 | 223 | 80 | 40 | 9 | 66 | 251 | 709 |
| android.location.Location.getLatitude | 41 | 222 | 79 | 40 | 9 | 65 | 251 | 707 |
| android.content.pm.PackageManager.queryIntentServices | 75 | 78 | 0 | 25 | 0 | 31 | 126 | 335 |
| java.net.HttpURLConnection.getInputStream | 13 | 104 | 11 | 11 | 20 | 9 | 59 | 227 |
| java.util.Locale.getCountry | 0 | 9 | 88 | 0 | 10 | 10 | 33 | 150 |
| android.content.pm.PackageManager.queryIntentActivities | 4 | 14 | 3 | 5 | 2 | 10 | 17 | 55 |
| android.accounts.AccountManager.getAccounts | 0 | 0 | 0 | 0 | 0 | 15 | 36 | 51 |
| android.content.pm.PackageManager.queryBroadcastReceivers | 0 | 18 | 0 | 7 | 2 | 4 | 9 | 40 |
| android.view.View findViewById | 0 | 30 | 0 | 0 | 0 | 0 | 0 | 30 |
| Others | 25 | 16 | 4 | 2 | 10 | 13 | 25 | 95 |
| **Total** | 364 | 948 | 269 | 161 | 66 | 240 | 1216 | 3264 |

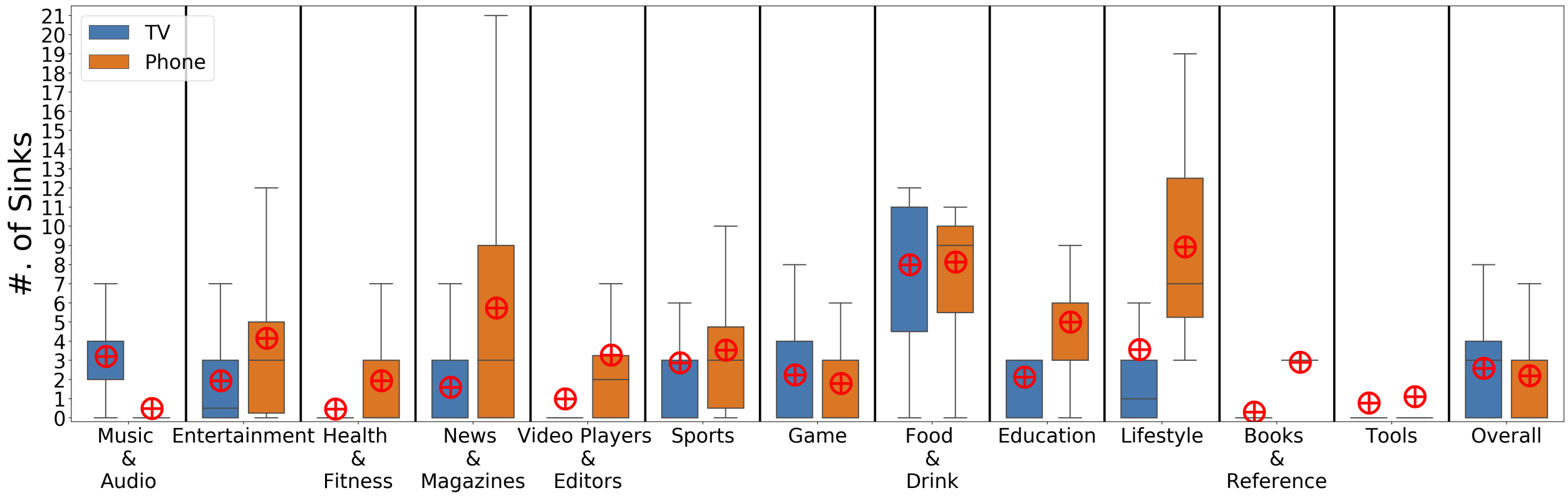

**Figure 11:** Distribution of FlowDroid Sinks in phone and TV apps for 755 Pairs with Different APKs

On average, Figure 10 presents that each phone app typically has more leaks than their TV counterpart in most categories, except for those in the Music & Audio and Game categories. However, the overall number of leaks revealed in phone apps (1,610 leaks) is less than that in their TV counterparts (1,950 leaks). It is caused by the fact that, as shown in the Music & Audio category in Figure 10, the number of pairs (i.e., 394) takes up more than half of the total pairs (i.e., 755), with 3.2 and 0.5 leaks detected in TV and phone apps, respectively.

As observed from Table 6, sensitive information is mostly leaked through the built-in Logcat functionality that developers typically employ for debugging. Users' data may leak from the logs, as attackers may obtain sensitive user information by inspecting the logs on the corresponding device. The second most frequently detected sink is the Bundle object. It is used to transport data between activities, processes, and configuration updates. A large amount of information contained in Bundle objects is exploitable by attackers in both phone/TV apps.

Overall, our results demonstrate that both security and privacy issues prevalently exist in TV apps. Also the TV and phone versions of the same app often encounter different kinds of security and privacy vulnerabilities. Our comparative experiments indicate that more security vulnerabilities are detected in phone applications while more privacy leaks are detected in TV applications. Therefore, developers and researchers should carefully examine the possible security vulnerabilities when developing apps for different devices.

**Table 6**
Distribution of Sinks detected with more than 20 occurrences by FlowDroid for 755 Pairs with Different APKs

| Sinks | Music & Audio | Entertainment | Food & Drink | Sports | News & Magazines | Lifestyle | Other | Total |
|---|---|---|---|---|---|---|---|---|
| **TV** | | | | | | | | |
| android.util.Log.d | 831 | 27 | 115 | 11 | 10 | 14 | 11 | 1019 |
| android.util.Log.w | 156 | 4 | 0 | 6 | 5 | 6 | 5 | 182 |
| android.os.Bundle.putString | 4 | 75 | 2 | 14 | 8 | 6 | 47 | 156 |
| android.util.Log.i | 125 | 3 | 0 | 2 | 0 | 0 | 1 | 131 |
| android.content.Context.bindService | 84 | 1 | 0 | 3 | 2 | 2 | 2 | 94 |
| java.lang.String.replace | 2 | 38 | 1 | 7 | 6 | 4 | 22 | 80 |
| java.io.ByteArrayOutputStream.write | 21 | 12 | 0 | 2 | 6 | 1 | 3 | 45 |
| android.content.Intent.setComponent | 18 | 5 | 0 | 6 | 2 | 4 | 10 | 45 |
| android.util.Log.v | 1 | 20 | 1 | 4 | 1 | 0 | 8 | 35 |
| Others | 22 | 51 | 1 | 20 | 13 | 13 | 43 | 163 |
| Total | 1264 | 236 | 120 | 75 | 53 | 50 | 152 | 1950 |
| **Phone** | | | | | | | | |
| android.os.Bundle.putString | 35 | 127 | 8 | 17 | 20 | 7 | 58 | 272 |
| android.util.Log.d | 31 | 56 | 51 | 15 | 33 | 24 | 30 | 240 |
| java.lang.String.replace | 5 | 65 | 11 | 8 | 12 | 4 | 38 | 143 |
| android.util.Log.v | 14 | 32 | 20 | 7 | 12 | 6 | 21 | 112 |
| android.content.Intent.setComponent | 15 | 23 | 1 | 11 | 11 | 6 | 15 | 82 |
| android.util.Log.w | 12 | 12 | 1 | 4 | 12 | 10 | 19 | 70 |
| android.content.Context.registerReceiver | 0 | 9 | 3 | 2 | 4 | 10 | 36 | 64 |
| android.os.Bundle.putParcelable | 4 | 17 | 5 | 1 | 1 | 4 | 21 | 53 |
| java.io.ByteArrayOutputStream.write | 11 | 9 | 0 | 3 | 9 | 2 | 8 | 42 |
| android.content.Context.startService | 14 | 10 | 0 | 5 | 6 | 0 | 4 | 39 |
| android.util.Log.i | 5 | 6 | 1 | 1 | 3 | 4 | 9 | 29 |
| java.io.ByteArrayOutputStream.write | 0 | 2 | 0 | 0 | 9 | 0 | 15 | 26 |
| android.util.Log.e | 9 | 8 | 0 | 2 | 3 | 2 | 2 | 26 |
| android.content.SharedPreferences.putString | 4 | 4 | 3 | 1 | 2 | 4 | 5 | 23 |
| android.content.Context.bindService | 3 | 8 | 0 | 2 | 3 | 2 | 4 | 22 |
| android.util.Log.e | 1 | 1 | 0 | 0 | 7 | 2 | 11 | 22 |
| Others | 30 | 108 | 18 | 13 | 40 | 38 | 98 | 345 |
| Total | 193 | 497 | 122 | 92 | 187 | 125 | 394 | 1610 |

**RQ 2: Summary**

- Slightly more security vulnerabilities are flagged in phone versions than TV versions, while more privacy leaks are detected in TV versions than phone versions.
- TV and phone versions of the same app often encounter different kinds of security vulnerabilities.
- The built-in Logcat functionality is the main leaking source for both phone/TV apps.
- Phone/TV apps in Music & Audio and Entertainment categories are more susceptible to security vulnerabilities, and also have more privacy leaks.

### 4.3. RQ3: User Interaction

We analyze the callback event management approaches in the app pairs. There are several methods for intercepting events generated by a user's interaction with Android apps. When it comes to events inside the UI, the best practice is to capture them from the individual View object that the user interacts with, which can be enabled by the View class. Various public callback methods for UI events are included in the different View classes that are used to compose layouts. (1) When the requested action is performed on the object, the Android framework executes these methods. For example, when a View (such as a Button) is touched, the *onTouchEvent()* function on that object is invoked. To intercept this, the class has to be extended to override this method, which is referred to as Event Handlers. (2) For easier managing interaction, Android introduces the Event Listener which is an interface where methods can be registered so that it can be triggered by the user interaction (e.g., *onTouch* method in the *android.view.View.OnTouchListener* interface can be registered by *setOnTouchListener* method in *View* class). (3) For the CLICK interaction, it also can be implemented by declaring the corresponding method in xml layout [17] as demonstrated in line six of the code snippet shown in Listing 1.

Event Handler and Event Listeners are used to handling the users' interaction when users trigger specific widgets. By analyzing the Event Listeners and Event Handlers in the apps, we want to discover how the logic of users' interaction differs between phone/TV apps. To this end, we decompile the bytecode to Java source code using AndroGuard [8],

```
<Button xmlns:android="http://schemas.android.com/apk/res/android"
    android:id="@+id/button_send"
    android:layout_width="wrap_content"
    android:layout_height="wrap_content"
    android:text="@string/button_send"
    android:onClick="sendMessage" />
```

Listing 1: Using *android:onClick* in *.xml* file.

**Table 7**
Distribution of the number of Top 10 Interaction Event for Pairs with Different APKs

| Interaction | Method | TV | Phone |
|---|---|---|---|
| CLICK | onClick, onClick attr in XML file | 103 | 220 |
| TOUCH | onTouch, onTouchEvent | 54 | 75 |
| KEY | onKeyDown, onKeyUp, onKeyLongPress, onKeyShortcut, onKeyMultiple, onKey | 48 | 43 |
| SCROLL | onScroll, onNestedScroll, onScrollChanged, onScrollChange | 31 | 42 |
| FLING | onFling, onNestedFling | 11 | 12 |
| HOVER | onHoverEvent, onHover | 10 | 11 |
| LONG CLICK | onLongClick | 4 | 5 |
| LONG PRESS | onLongPress | 3 | 3 |
| DOWN | onDown | 3 | 3 |
| TRACK BALL | onTrackballEvent | 2 | 1 |

and search for the presence of Event Listeners and Event Handlers of user input provided in official Android documentation [15]. In addition, for the CLICK interaction, we further count the occurrence of *onClick* attribute that appeared in the layout (.xml) files in each app.

The top ten user interactions in phone/TV app pairs are shown in Table 7. As indicated, the top 10 user interaction types are the same in TV and phone apps. The most involved user interaction event on both phones and TVs is *CLICK*. On average, each phone app has 220 CLICK-related inputs, more than twice the number seen in a TV app (i.e., 103).

In general, phone apps get more user input than TV apps. The exceptions are the KEY-related input, which is declared on average 43 times in smartphone apps and 48 times in TV apps, and the input related to TRACK BALL, which is on average declared twice in each smartphone app and only once in each TV app.

As mentioned before, a View is a widget generally used to display something like Buttons, ListViews, etc. To make these Views look well-organized, Android introduces the *Activity* class, where the View component can be placed in the UI by using *setContentView(View)*. Activities are primarily shown to the user as full-screen windows; however, they may alternatively be displayed as floating windows (through a theme with the *R.attr.windowIsFloating* property set) in the Multi-Window mode, or embedded inside other windows. As a result, the number of Activities in an Android app may reflect the complexity of the UI layout to some extent. Figure 12 compares the number of Activities in phone/TV pairs for each category. As shown, the average number of Activities (i.e., 24) in smartphone apps is more than twice the number (i.e., 10) in TV apps. One possible reason may be that the larger screen size on TV could contain more content and so that fewer UI screens (i.e., activities) are needed.

Instead of utilizing touch screens, TV users must rely on peripherals, such as a remote controller, to engage with TVs, which is distinct from how users interact with their phones. However, our empirical results show that the interaction events in the apps on both devices are comparable. The evidence above demonstrated the possibility of a reuse of code for user interaction in the phone apps and their TV versions.

**RQ 3: Summary**

- TV apps encounter fewer user interactions than their phone versions, but the type of user interaction events are similar between phone/TV apps.
- CLICK-related operations are most commonly used in both phone and TV apps for user interaction. TV apps have significantly fewer CLICK events than phone apps.
- TV apps significantly reduced the number of screens (i.e., Activities) compared with their phone versions.

## 5. Discussion

This section discusses implications of this study and threats to validity of our results.

**Implication.** The results of the study highlight a number of concerns and possibilities for the research community and practitioners.

*Possibility of automated migration.* As unveiled in our empirical findings, around 44% to 50% of the code in TV apps is similar with their smartphone counterparts, and 45% of resources are shared between them. Given the disparity in market share between phone/TV apps and the popularity of smart TVs, there is a high demand for the TV versions of smartphone apps. Developing an automated method for migrating smartphone apps to TV would alleviate developers' engineering efforts and contribute to the Android ecosystem. Our empirical research can serve as a starting point for the development of such an automated migration tool.

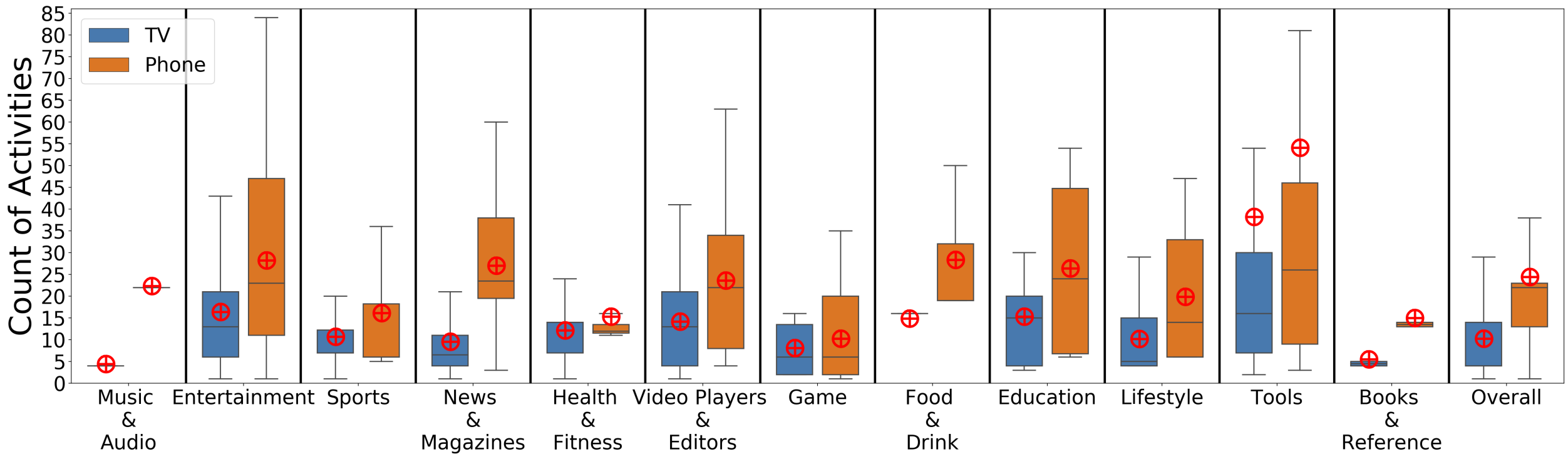

**Figure 12:** Distribution of screens in phone and TV apps for Pairs with Different APKs

*TV apps could introduce even more serious security/privacy issues compared with their phone counterparts.* It has been commonly assumed that the TV version of an app is usually a simplification of the smartphone version. Indeed, through analyzing the collected phone/TV app pairs, we found that less feature and interaction event contained in TV apps. However, the security and privacy issues in certain categories of TV apps are quite prominent. These experimental results confirm that security/privacy issues could have been overlooked in the TV version of the Android apps. Developers hence need to pay more attention to secure TV apps, and TV app specific analysis tools are in demand to better facilitate this requirement.

**Threats To Validity.** Our study is conducted based on a limited number of phone/TV app pairs, which may not be representative of the whole Android TV ecosystem. As mentioned in Section 3, we simulate the corresponding devices when querying the Google Play store for downloading the apps. The search results are highly affected by device settings and capabilities (e.g., device types, OS versions, and hardware architectures). Our simulated devices may not be compatible with all apps, resulting in missing some apps in our collection. However, we simulated the devices with the latest OS versions (i.e., API level 30), which could reduce the number of missed apps in our collected dataset. Also, our dataset was collected based on one of the most comprehensive datasets, AndroZoo, which demonstrates the landscape of our dataset.

## 6. Related Work

Previous research has conducted extensive analyses of the metadata and code of phone apps on a broad scale. However, the features of TV apps have received little attention. To the best of our knowledge, our work is the first comparative study between an app's phone version and a TV version. We present and discuss briefly related studies: (1) Large-scale Android application analysis, (2) Android security.

**Large-scale Android application analysis.** Previous studies have extensively explored Android apps from different aspects. Numerous studies have been conducted to analyze the metadata of phone apps [42, 19, 26]. Wang *et al.* [42] conducted an extensive study on 6 million Android apps downloaded from 17 different app markets to understand catalog similarity across app stores. This large-scale comparative study discovered that Chinese third-party markets present a higher prevalence of fake, cloned, and malicious applications than Google Play. Chen *et al.* [6] collected 223 pairs of Android Smartphone and Smartwatch apps and examined them from both non-code and code aspects to understand the relationship between them. At the same time, there are several works that have studied the Android TV ecosystem. Liu *et al.* [30] empirically examined the security issues of 3,163 Android TV applications. Tileria *et al.* [41] presented a behavior analysis on 4.5k Android TV applications and their results proved that TV apps face serious security and privacy issues.

However, prior studies mainly focused on Android smartphone apps analysis. Android TV apps have not yet been thoroughly investigated. Although the prior works have attempted to explore the Android TV ecosystems, the similarity between smartphone and TV applications are still unknown. In contrast to previous work, we empirically examine the similarity and differences between Android TV applications and their smartphone counterparts.

**Android security.** Android is one of the most popular open-source operating systems, particularly on smartphones [31]. With this widespread usage, an increasing number of Android applications are developed. This rising popularity, on the other hand, draws malware developers and exposes Android users to a growing number of security threats [10]. Malware app developers exploit platform vulnerabilities and steal private user data for profit. The empirical study conducted by Linares *et al.* [28] examined 660 Android-related vulnerabilities such as memory corruption and data handling and discovered that Android vulnerabilities survive for a long period of time (at least 724 days, on average) in the codebase. Thus, vulnerability detection and analysis have garnered considerable attention from the academic community, and numerous relevant research works have been proposed to improve software security [36, 43, 47, 7, 21]. For example, Zhan *et al.* [46] proposed an obfuscation-resilient third-party library detection tool that can precisely locate vulnerable third-party libraries used in Android applications. Apart from that, Android malware detection and defense have also attracted increasing research attention [38]. To better analyze apps' features and behaviors, industry and academia have proposed a variety of security and privacy analyzers for Android, including static [5, 22], dynamic [39, 32] and hybrid [29]. Zhou *et al.* [49] developed a rule-based static malware detection system (i.e., permission-based behavioral footprinting and heuristics-based filtering) to detect new malware samples. In recent years, machine learning and deep learning techniques have been widely introduced to detect Android malware and demonstrated promising performance [4, 33, 34, 48]. For instance, Mclaughlin *et al.* [34] employed a deep convolutional neural network (CNN) to construct an Android malware detector and their experiments demonstrated that the proposed model can achieve a 98% detection accuracy.

To the best of our knowledge, previous works mainly target the security of smartphone apps. We investigate the security and privacy risks in Android phone/TV pairs. As our future work, we plan to go beyond this work by inventing better approaches for improving the security and reliability of Android TV apps.

## 7. Conclusion

In this work, we present the first study towards understanding and characterizing Android phone/TV app pairs. We gathered and analyzed 973 Android phone/TV pairs with distinct APKs. We specifically examined the similarity of gathered app pairs from both non-code and code perspectives. Our experimental investigation finds that on average, 43% of resource files (e.g., images, UI layouts, etc.) and 50% of code are reused between Android phone/TV app pairs, which suggests that apps can be prioritized for migration from smartphones to smart TVs. We also analyze the interaction between phone/TV pairs and find that events triggered by user interaction are similar between phone/TV pairs. In addition, we investigated how security and privacy issues are involved in phone/TV pairs, our findings revealed that the security and privacy issues in TV apps have been overlooked. Specifically, TV apps in certain categories (e.g., Music & Audio, Entertainment) tend to include more security and privacy issues than their phone counterparts. Therefore, we encourage our community to pay more attention to Android TV apps, especially from the security perspective.